\documentclass[prl, superscriptaddress, twocolumn, showpacs, floatfix, nobalancelastpage]{revtex4-1}
\usepackage{amsfonts}
\usepackage{amssymb}
\usepackage[dvips]{graphicx}
\usepackage{amsmath}
\usepackage{color}

\newcommand{\ket}[1]    {| #1 \rangle}

\newcommand{\ii}         {\mathrm{i}}

\renewcommand{\t}[1]{\textrm{#1}}
\newcommand{\mat}[2]{
\left(\begin{array}{#1}
#2
\end{array}
\right)}
\begin{document}
\title{Fundamental quantum interferometry bound for the squeezed-light-enhanced gravitational-wave detector GEO\,600}
\author{Rafa{\l} Demkowicz-Dobrza{\'n}ski}
\affiliation{Faculty of Physics, University of Warsaw, ul. Ho\.{z}a 69, PL-00-681 Warszawa, Poland}
\author{Konrad Banaszek}
\affiliation{Faculty of Physics, University of Warsaw, ul. Ho\.{z}a 69, PL-00-681 Warszawa, Poland}
\author{Roman Schnabel}
\affiliation{Institut f\"ur Gravitationsphysik, Leibniz Universit\"at Hannover, Max-Planck-Institut f\"ur Gravitationsphysik, Callinstra{\ss}e 38, D-30167 Hannover, Germany}

\begin{abstract}
The fundamental quantum interferometry bound limits the sensitivity of an interferometer for a given total rate of photons and for a given decoherence rate inside the measurement device.
We theoretically show that the recently reported quantum-noise limited sensitivity of the squeezed-light-enhanced gravitational-wave detector GEO\,600 is exceedingly close to this bound, given the present amount of optical loss. Furthermore, our result proves that the employed combination of a bright coherent state and a squeezed vacuum state is generally the optimum practical approach for phase estimation with high precision on absolute scales. Based on our analysis we conclude that neither the application of Fock states nor N00N states or any other sophisticated nonclassical quantum states would have yielded an appreciably higher quantum-noise limited sensitivity.
\end{abstract}

\maketitle

Direct detection of gravitational waves (GWs) is one of the most challenging tasks in contemporary experimental physics.
Over the recent years, advancements in the design and in the practical realization of GW detectors have led to significant reduction of technical noise. Even for kilowatts of circulating light powers, the performance of the devices has approached the precision limits imposed by the laws of physics themselves.
Modern day GW detectors are kilometre-scale laser interferometers in which suspended mirrors play the role of test masses of space-time curvature.
Displacements of the mirrors induce variation in the light power leaving the output port of the interferometer.
Thanks to the high light power circulating inside the interferometer, tiny changes in the mirrors' relative positions
lead to a measurable change in the number of output photons---opening up prospects of detecting GWs where
the typical relative mirror motion amplitudes are expected to be below the size of the proton.
The dominating noise source in GW detectors at signal frequencies above a couple of hundreds of hertz is
the photon shot noise \cite{Pitkin2011, Cella2011}.
Standard laser light is well described by a coherent state, which implies that the number of photons $n$ registered at the output port
fluctuates according to the Poissonian statistics as $n = \bar{n} \pm \sqrt{\bar{n}}$,
requiring a signal to provide at least order of $\sqrt{\bar{n}}$ change in the mean registered photon number $\bar{n}$
to provide a signal-to-noise ratio of one.

It has been proposed already in the early 1980s that the use of  a squeezed vacuum state may lead to an improved performance of GW detectors without the need for increasing the number of photons \cite{Caves1981, Bondurant1984}.
The proposal has recently found its full-scale realization in the GEO\,600 GW detector where sub-shot noise sensitivities have been demonstrated in a GW detection operating mode \cite{LIGO2011}. More recently an operation over several months has been reported \cite{Grote2013b}, as well as a successful test in one of the LIGO detectors \cite{LIGO2013}.
\begin{figure}[t]
  \includegraphics[width=0.9\columnwidth]{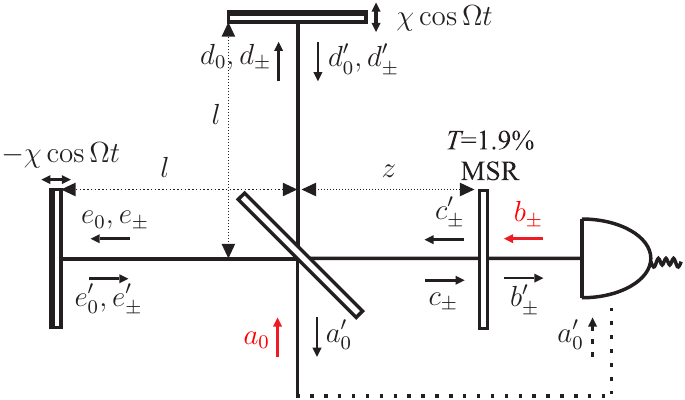}\\
  \caption{A simplified model of the GEO\,600 interferometer operating at a dark fringe.
  $a_0, d_0,e_0$ represent anihilation operators of the corresponding modes at
 the  carrier frequency $\omega_0$, while $b_{\pm}, c_{\pm}, d_{\pm}, e_{\pm}$
 represent annihilation operators of sideband modes $\omega_0 \pm \Omega$. Input modes are marked in red.
 The signal recycling mirror (MSR) with power transmissivity of 1.9\% is placed in the signal output port and is responsible for the frequency dependent amplification of the signal amplitudes. The distances to the two end mirrors oscillate at frequency $\Omega$
 with the relative phase shift $\pi$ and amplitude $\chi$.
  }\label{fig:geo600}
\end{figure}
Other approaches based on the use of Fock and the N00N states \cite{Holland1993, Dowling1998, Mitchell2004, Afek2010}
may also lead to an improved sensitivity but the improvement is strongly bounded by the effects of optical loss
\cite{Dorner2008, Kolodynski2010, Knysh2011, Escher2011, Demkowicz2012}.

Here we prove that the quantum enhancement of sensitivity based on
the interferometric scheme combining coherent states and squeezed vacuum (CSV) as reported in  \cite{LIGO2011}
was close to the fundamental quantum bound under given energy constraints and optical loss levels.
Only a small increase of the squeezing factor would have virtually met this bound.

We consider a simplified model of the GEO\,600 interferometer consisting of a Michelson interferometer with a signal-recycling cavity \cite{GEO6002010}, as depicted in Fig.~\ref{fig:geo600}.  Our model omits the power recycling cavity of GEO\,600, and instead assumes that the actual light power at the beam splitter $P$ is achieved by sending in a light field of correspondingly higher power.
Besides the light power the important experimental parameter of the system is its optical power input-output transfer coefficient $\eta$, where $(1\!-\!\eta)$ represents
the cumulative effect of photon scattering, absorption,
mode mismatch, and photo detection efficiencies.
For the squeezed vacuum state sent into the asymmetric port of the interferometer in the GEO\,600 setup $\eta$ was measured to be $0.62$ \cite{LIGO2011}.
Phase noise decoherence effect on the squeezed state \cite{Franzen2006} was considered to be negligible.
Also the effect of measurement back-action due to photon radiation pressure \cite{Caves1980} is currently negligible in GEO\,600 and is not
 included in our analysis. However, back-action can in principle be avoided \cite{Kimble2001} and thus does not limit the scope of our analysis.

Since a general interferometric scheme might involve arbitrary quantum states of light being sent into \emph{both} input ports, we make for simplicity the conservative assumption that the same overall loss would be experienced for both paths.
In reality the input-output transfer coefficient of the GEO\,600 bright port is even lower than $0.62$ due to the high finesse of the power-recycling cavity being almost impedance matched.
Without loss of generality we further assume that in the absence of gravitational waves
the interferometer output port is at a dark fringe and its arms have perfectly equal lengths $l$. Additional phase shifts
could always be included in the general forms of the input state or the measurement scheme and as such would have no impact on fundamentally achievable precision.
Oscillations of the differential distance of the end mirrors
at a frequency $\Omega$ induce  an exchange of light power between the carrier light's optical frequency $\omega_0$ and the sideband modes $\omega_0 \pm \Omega$,
which effectively transfers light between the two output ports of the interferometer.
One can regard the system as a linear transformation between the three input modes
$a_0, b_\pm$ to the three output modes $a_0^\prime, b^\prime_{\pm}$ as depicted in Fig.~\ref{fig:geo600}, where the three indices $0, \pm$ refer to the frequencies $\omega_0, \omega_0 \pm \Omega$.

In our analysis we apply no restriction to the nature of the quantum states to be sent into the apparatus, nor do we restrict ourselves to specific measurements performed at the two output modes.
The performance of the detection is
quantified by a signal-to-noise-ratio of unity for the measurement of a gravitational wave strain $h = 2\chi/l$, where $\chi$ is the amplitude of the mirror oscillations in each arm, as shown in Fig.~\ref{fig:geo600}.

We first consider three modes $d_0, d_{\pm}$ impinging on the top mirror in Fig.~\ref{fig:geo600}.
Assuming the amplitude of the latter's oscillations much smaller than the light wavelength, $\chi \ll \lambda_0=2\pi c/\omega_0$, and neglecting higher order sideband modes $\omega_0 \pm n \Omega$ for $n\geq 2$,
we can approximate the input-output relation to first order in $\epsilon = h l \omega_0/ 2 c $ as
\begin{equation}
\mat{c}{
d^\prime_0 \\
d^\prime_+ \\
d^\prime_- } =
\mat{ccc}{
1 & \ii\epsilon &  \ii \epsilon  \\
\ii \epsilon & 1 & 0 \\
\ii \epsilon & 0 & 1
}
\mat{c}{
d_0 \\
d_+ \\
d_- }.
\end{equation}
An analogous relation holds for $e_0, e_\pm$ modes with $\epsilon$ replaced by $-\epsilon$
to account for the antisymmetric nature of the GW signal.

In the GEO\,600 experiment the signal-recycling cavity was tuned to the central frequency $\omega_0$. We realize this configuration by setting $z=0$ (see Fig.~\ref{fig:geo600}) and by setting $\omega_0 l/c$ to a multiple of $\pi$.
Up to $\epsilon$-independent phase factors, which are irrelevant for the discussion of
sensitivity, the input-output relations have the form
\begin{equation}
\label{eq:u}
\!\mat{c}{
a_0^{\prime}\\
b_+^{\prime}\\
b_-^{\prime}
}\!=
U \!
\mat{c}{
a_0\\
b_+\\
b_-
}\!,
\
U\!=\!\mat{ccc}{
1 & g \epsilon & g \epsilon \\
- g\epsilon & 1 & 0 \\
- g \epsilon & 0 & 1
},
\end{equation}
where
$$g = \sqrt{\frac{T}{2-T-2\sqrt{1-T} \cos [2 \Omega l /c]}}$$ is the amplification factor due to the presence of the signal-recycling mirror with power transmissivity $T$.
\begin{figure}[t]
  \includegraphics[width=0.9\columnwidth]{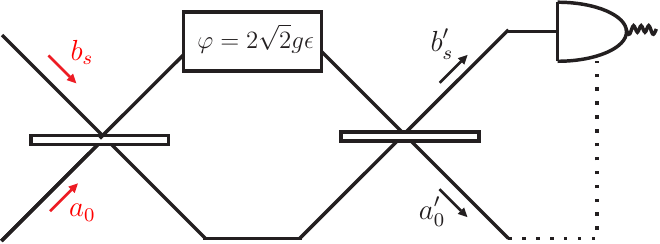}\\
  \caption{Model of the GEO\,600 interferometer reduced to an equivalent, in terms of sensitivity, two-mode Mach-Zehnder interferometer
  with relative phase delay  $\varphi=2\sqrt{2}g \epsilon$.   $b_s$ represents a symmetric combination of the sideband modes.
  The possibility of a more general measurement is illustrated with the dotted line.   }\label{fig:machzehnder}
\end{figure}

Inspecting Eq.~\eqref{eq:u} we notice that the effective mode coupling occurs between the central $a_0$
and the \emph{symmetrized} sideband mode $b_s = (b_- + b_+)/\sqrt{2}$ reads
\begin{equation}
\mat{c}{
a_0^\prime\\
b_s^\prime
} =
\mat{cc}
{
1 & \sqrt{2}g \epsilon \\
-\sqrt{2}g \epsilon & 1
}
\mat{c}{
a_0 \\
b_s
} \, ,
\end{equation}
leaving the antisymmetric mode $b_a = (b_- - b_+)/\sqrt{2}$ intact.
We may look at this effective evolution in terms of an equivalent Mach-Zehnder interferometer with small relative phase shift
$\varphi=2\sqrt{2} g \epsilon$, where the pairs of input and output modes
are represented by $a_0,b_s$ and $a^\prime_0,b_s^\prime$, respectively, as shown in Fig.~\ref{fig:machzehnder}.

We first recall the theoretical model that is valid for an interferometer with modes $a_0$ and $b_s$ being in a coherent state $\ket{\alpha}$ and a squeezed vacuum state $\ket{r}$, respectively. Here, $e^{-2r}$ is the squeezing factor of the quadrature variance.
In the experimentally relevant limit when the classical beam is much stronger than the squeezed
one ($|\alpha| \gg \t{sh}^2 r$) and when the power transmission $\eta$ is identical for both arms,
the phase estimation uncertainty obtained by simply measuring the output light power (being proportional to $b_{s}^{\prime\dagger} b_{s}^\prime$) is approximately given by  \cite{Caves1981}
\begin{equation}
\label{eq:phase_squeezed}
\Delta \varphi \approx \sqrt{\frac{1- \eta+ \eta e^{-2 r} }{\eta |\alpha|^2}} \, . %\overset{r \gg 1}{\approx} \sqrt{\frac{1 - \eta}{\eta |\alpha|}}.
\end{equation}
The enumerator can be interpreted as a combination of contributions from the squeezed and the vacuum quadrature
variances with respective weights $\eta$ and $1-\eta$ defined by the setup losses \cite{Banaszek1997}.

Inserting into Eq.~\eqref{eq:phase_squeezed} the relation $h = \varphi c/\sqrt{2} g l $ and
the mean photon number produced in unit time by the light source of power $P$, $|\alpha|^2 =
P/(\hbar \omega_0)$, we arrive at the (frequency domain) single-sided strain-normalized noise spectral density
\begin{equation}
\label{eq:squeezed}
\Delta h = \frac{1}{l g} \sqrt{\frac{c \hbar \lambda_0}{4 \pi P}} \sqrt{\frac{1-\eta +\eta e^{-2r}}{\eta}} \, .
\end{equation}
The above formula is valid for an interferometer whose decoherence is dominated by optical loss being independent of the input port.
To apply Eq.~\eqref{eq:squeezed}  to the actual data as presented in Ref.~\cite{LIGO2011} we have to re-scale the measured light power at the beam splitter of $\tilde{P} \approx 2.7\t{\,kW}$, which already experienced some optical loss.

We now set $\eta=0.62$, which was measured for the squeezed vacuum state
being reflected from the interferometer's signal output port \cite{LIGO2011}. A reasonable Ansatz is to decompose the efficiency in an in-coupling and out-coupling efficiency ($\eta=\eta_{in}\cdot\eta_{out}$). Since imperfect mode-matching mainly affects the input efficiency we conservatively estimate $\eta_{in}$ to be about 0.73. Using Eq.~\eqref{eq:squeezed} and the correspondingly higher value for the circulating power of $P = \tilde{P} / \eta_{in} = 3.7$\,kW corresponds to an out-coupling efficiency of the signal sidebands of 0.85 being a reasonable value for GEO\,600 \cite{Grote2013}. We estimate that the above approximation leads to an error well within the measurement error for $\tilde{P} \approx 2.7\t{\,kW}$ of about 20\% \cite{Leong2012}.

\begin{figure}[t]
  \includegraphics[width=0.9\columnwidth]{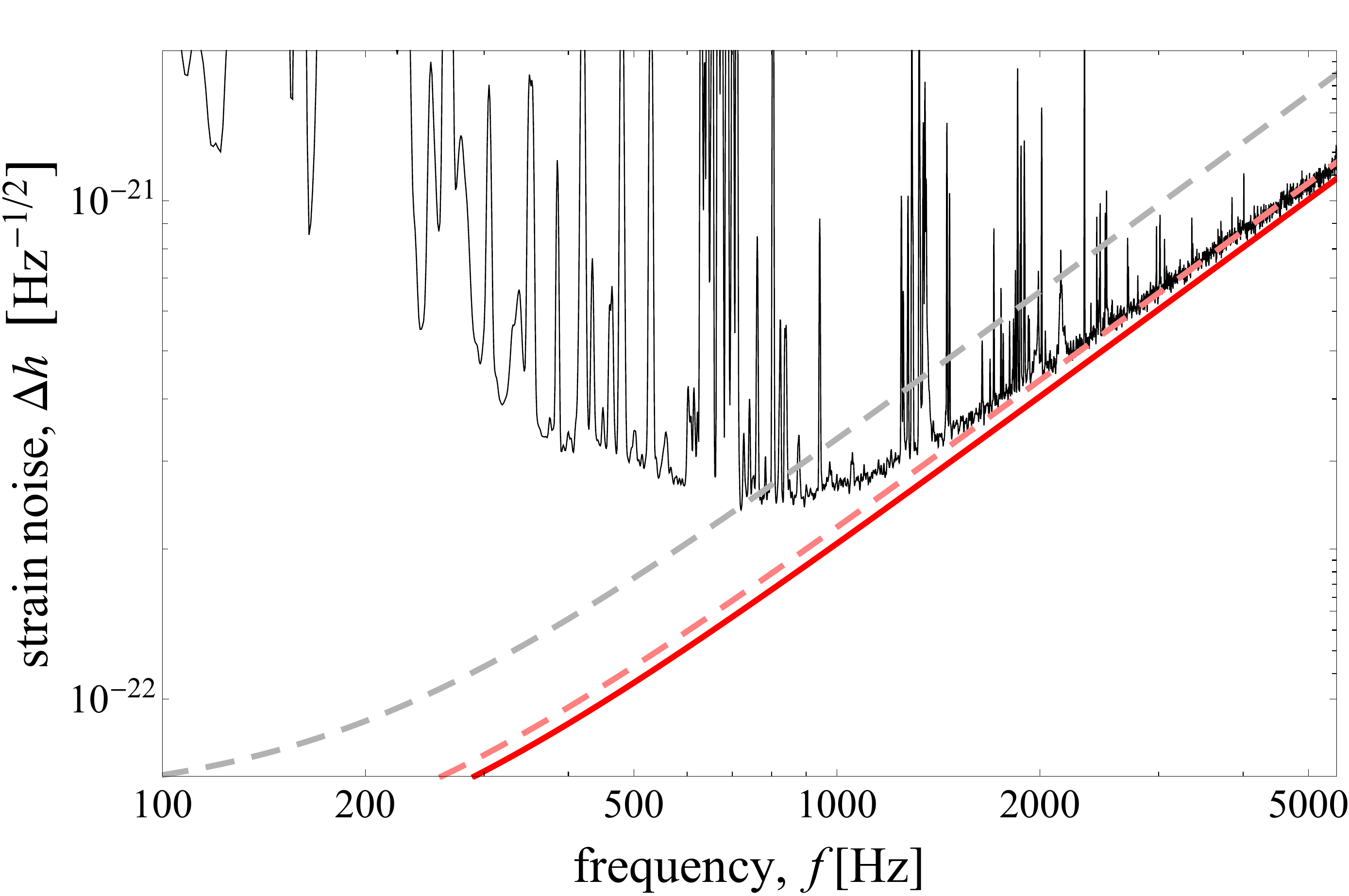}\\
  \caption{GEO600 noise spectral density normalized to strain ($\Delta h$, black) \cite{LIGO2011} and predictions of our simplified theoretical model based on given values for the circulating light power and for the given optical loss: no vacuum squeezing (dashed gray), 10\,dB vacuum squeezing (dashed red,  Eq.~\eqref{eq:squeezed}) as realized in \cite{LIGO2011}, and
  16\,dB vacuum squeezing coinciding with the fundamental quantum enhancement bound (solid red, Eq.~\eqref{eq:fundamental}).
Note that the injection of 16\,dB vacuum squeezing is in reach of current technology, since 12.7\,dB were already observed, even with imperfect photo detectors \cite{Eberle2010}. Measurement data courtesy of The LIGO Scientific Collaboration.}
  \label{fig:resultsmodel}
\end{figure}

Fig.~\ref{fig:resultsmodel} (red dashed line) shows our model described above in comparison to the experimental data presented in \cite{LIGO2011}. Both traces match rather well above 1500\,Hz, where the experimental data is clearly dominated by quantum noise. Our model (Eq.~\eqref{eq:squeezed}) uses the following
experimental parameters \cite{LIGO2011}: $\lambda_0=1064 \t{\,nm}$, $l=1200 \t{\,m}$, $P=3.7 \t{\,kW}$, $\eta=0.62$, $T=1.9\%$, $e^{-2r}\!=0.1$.

The main mathematical concept behind the derivation of the fundamental quantum interferometry bound is the
\emph{quantum Fisher information} (QFI) \cite{Helstrom1976, Braunstein1994}.
Let  $\rho$ be the input state sent into the interferometer while $\Lambda_\varphi$ represent the total action of the interferometer including  the phase delay $\varphi$ and loss. QFI calculated on the output state $F(\Lambda_\varphi(\rho))$, which for brevity we denote as $F(\rho)$, provides a limit on the achievable phase estimation sensitivity via the Cram\'{e}r-Rao bound
\begin{equation}
\Delta \varphi \geq \frac{1}{\sqrt{F(\rho)}}\, .
\label{eq:cr}
\end{equation}

In recent years the influence of non-zero optical loss was considered \cite{Dorner2008, Kolodynski2010, Knysh2011, Escher2011, Demkowicz2012}.
For a generic two-mode input state $\rho_N$ with a precisely defined total photon number $N$ the limit on phase sensitivity is \cite{Knysh2011, Escher2011, Demkowicz2012}
\begin{equation}
\label{eq:boundfish}
\max_{\rho_N} F(\rho_N) \leq  \frac{N \eta}{1-\eta} \, , \, \rm{with} \,\; \eta < 1 \, .
\end{equation}

Here we present a generalized bound also applicable to states having an uncertain photon number such as coherent states and squeezed states as required for the setup investigated here. Let us also observe that in our setup no additional reference beams are involved. Consequently, any kind of measurement on the output beams is necessarily
a photon number measurement and hence coherences between different \emph{total photon number subspaces} of the two-mode density matrix are not observable
\cite{Molmer1997, Bartlett2006}. If, on the other hand, one assumed additional phase reference beams, the whole problem of phase estimation would need to be reformulated by specifing which relative phase is actually being estimated and which photons are included in the total power budget \cite{Jarzyna2012}.
Assuming that no additional reference beams are present, we can equivalently write any kind of a two-mode state with \emph{uncertain} photon numbers,
e.g.\ $\ket{\alpha}\otimes \ket{r}$,
  as an incoherent mixture of states having a \emph{certain} photon number in both modes
 \begin{equation}
  \rho_{\bar{N}} \equiv \bigoplus_{N=0}^\infty p_N \rho_N \, ,
 \end{equation}
where $p_N$ is the probability of projecting the state $\rho_{\bar{N}}$ onto the $N$-photon subspace, and $\rho_N$ is the normalized
conditional density matrix in the $N$-photon subspace.

The maximization of $F$ over states $\rho_{\bar{N}}$ with an indefinite photon number but the mean value fixed to $\bar{N}$
may therefore be carried out by taking normalized density matrices in $N$-photon subspaces
written as
\begin{equation}
F(\rho_{\bar{N}}) = F\left(\bigoplus_{N=0}^\infty p_N \rho_N\right) \, ,
\end{equation}
with a constraint $\sum_{N=0}^\infty p_N N = \bar{N}$. Thanks to the convexity of $F$ \cite{Fujiwara2001, Petz1996} we get
(for $\eta < 1$)
\begin{equation}
F(\rho_{\bar{N}}) \leq \sum_{N=0}^\infty p_N F(\rho_N) \leq \sum_{N=0}^\infty p_N \frac{N \eta}{1-\eta} = \frac{\bar{N}\eta}{1-\eta} \, ,
\end{equation}
where we have applied Eq.~\eqref{eq:boundfish} to each of the photon number subspace separately. This leads to the bound on the ultimate phase sensitivity in the form
\begin{equation}
\label{Eq:Deltaphiindef}
\Delta \varphi \geq \sqrt{\frac{1-\eta}{\eta \bar{N}}} \, ,
\end{equation}
where compared to previous results the average $\bar{N}$ appears in lieu of the definite number of photons.
Let us point out that the above derivation was possible only thanks to the linearity of the
bound on the QFI in $N$ and would break down in the decoherence-free case, where QFI grows as $N^2$. In general,
taking a sensitivity bound derived under an assumption of a fixed number of $N$ photons and just
replacing $N$ with an average $\bar{N}$ for arbitrary states
is illegitimate and may lead to apparently contradicting statements on the possibility of beating
the so called Heisenberg limit \cite{Hyllus2010, Anisimov2010, Rivas2012, Giovanetti2012}.

 \begin{figure}[t]
  \includegraphics[width=0.8\columnwidth]{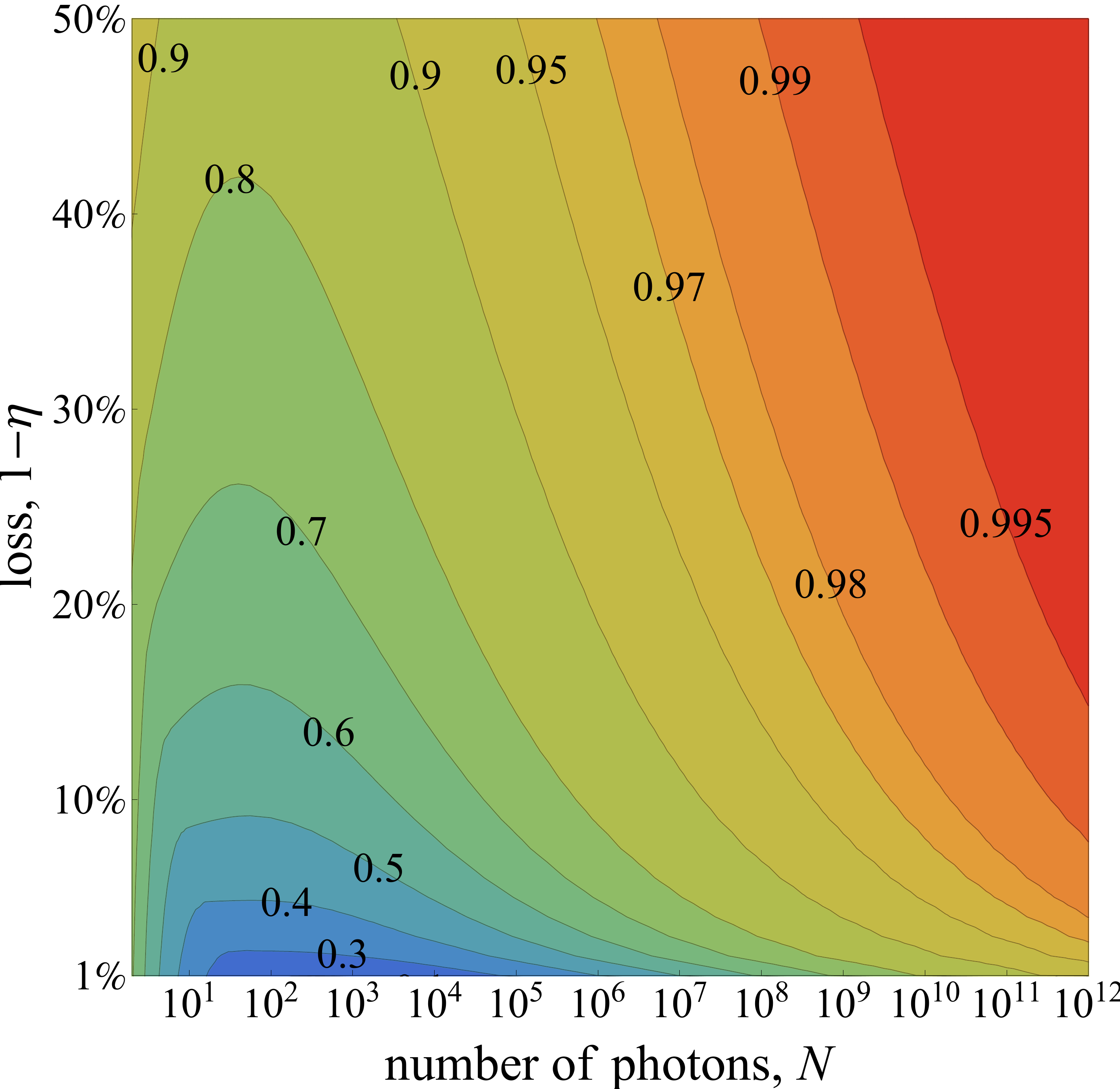}\\
  \caption{Optimality of the `coherent state\,--\,squeezed vacuum state' (CSV) scheme quantified as the ratio of precisions achievable with the optimal $N$ photon state and the CSV strategy with the same mean number of photons and optimal squeezing.
  Note that in GEO\,600, the effective number of photons used per $1\t{\,s}$ was approximately $2 \cdot 10^{22}$. For such a high photon number the CSV strategy would achieve 99.9996\% of the optimum quantum strategy's sensitivity for a loss of 38\%.}
  \label{fig:squeezing}
\end{figure}

Making use of Eq.~\eqref{Eq:Deltaphiindef} we are therefore entitled to write the fundamental bound on interferometer strain sensitivity as
\begin{equation}
\label{eq:fundamental}
\Delta h = \frac{1}{l g}\sqrt{\frac{c \hbar \lambda_0}{4 \pi P}} \sqrt{\frac{1-\eta}{\eta}} \, ,
\end{equation}
which is depicted as the solid red line in Fig.~\ref{fig:resultsmodel}.
Most interestingly, virtually the same spectral density is provided by Eq.~\eqref{eq:squeezed} when using 16\,dB ($e^{-2r}\!=0.025$) of squeezing instead of 10\,dB.

We conclude that there is no need for an alternative to the CSV interferometric strategy in the regime of
high light powers. The actual meaning of ``high'', however, depends on the loss level $1-\eta$.
The higher the losses the sooner the CSV strategy becomes optimal.
Fig.~\ref{fig:squeezing} is a contour plot of the ratio of precision achievable with the most general optimal $N$ photon states of light
and the precision achievable with the CSV strategy, with $\bar{N}=|\alpha|^2 + \t{sh}^2 r =N$, as a function of $N$ and the loss level $1-\eta$.
Optimal precision achievable with general states has been found by a direct numerical optimization of QFI up to $N=100$ \cite{Dorner2008}. For larger
$N$ we found a rather accurate extrapolation formula
\begin{equation}
\Delta \varphi = \sqrt{\frac{1-\eta}{\eta N} \left[1+ \frac{1}{\sqrt{N}}\left(a + \frac{b}{N} + \frac{c}{N^2}\right) \right]} \, ,
\end{equation}
where parameters $a$, $b$, and $c$ are fitted using low $N$ data. This model
is consistent with the asymptotic formulas derived in \cite{Knysh2011}.
For currently realistic loss levels of $1-\eta \approx 30\%$, already $N=2 \cdot 10^{9}$ guarantees less than 1\% deviation of the CSV scheme from  the optimal strategy. However, future GW detectors will use photon rates of the order 10$^{24}$. Then even loss levels below 1\% will still keep the CSV scheme within $1\%$ of the fundamental quantum interferometry bound.

Even though the presence of decoherence diminishes the potential gains offered by quantum metrology,
the fact that rather realistic strategies based on squeezed states allow to make the most of quantum enhancement is
highly encouraging. Analogous claims might also be made for quantum-enhanced atomic
clock calibration in the presence of dephasing, where theoretical results indicate that
the precision of Ramsey interferometry with spin-squeezed states is close to the optimal one
in the asymptotic regime of a large number of atoms \cite{Huelga1997, Orgikh2001, Escher2011}.
~\\
\newpage
We thank Wojciech Wasilewski, Micha{\l} Karpi{\'n}ski, and Dmitry Simakov for many fruitful discussions as well as Hartmut Grote and Jonathan Leong for discussions on the calibration of GEO\,600 parameters.
This research was supported by Polish NCBiR under the ERA-NET CHIST-ERA project QUASAR
Foundation for Polish Science TEAM project co-financed by the EU European Regional
Development Fund and FP7 IP project Q-ESSENCE.

\bibliography{geo600optimality}
\end{document}